\documentclass[lettersize,journal]{IEEEtran}
\IEEEoverridecommandlockouts
\usepackage{cite}
\usepackage[tbtags]{amsmath}
\usepackage{amssymb,amsfonts}
\usepackage{algorithmic}
\usepackage{graphicx}
\usepackage{textcomp}
\usepackage{xcolor}
\usepackage{caption}
\usepackage{commath}

\def\BibTeX{{\rm B\kern-.05em{\sc i\kern-.025em b}\kern-.08em
    T\kern-.1667em\lower.7ex\hbox{E}\kern-.125emX}}

\newtheorem{theorem}{\textbf{Theorem}}
\newtheorem{corollary}{\textbf{Corollary}}

\begin{document}

\title{A Bayesian Approach to Characterize Unknown Interference Power in Wireless Networks}

\author{Mahmoud Zaher, Emil Björnson, \IEEEmembership{Fellow, IEEE,} and Marina Petrova, \IEEEmembership{Member, IEEE.}%
\thanks{M. Zaher and E. Björnson are with the Division of Communication Systems, KTH Royal Institute of Technology, 164 40 Stockholm, Sweden (e-mail: mahmoudz@kth.se; emilbjo@kth.se).}

\thanks{M. Petrova is with the Division of Communication Systems, KTH
Royal Institute of Technology, 164 40 Stockholm, Sweden, and also with the Mobile Communications and Computing Group, RWTH Aachen University, 52062 Aachen, Germany (e-mail: petrovam@kth.se).}
\thanks{This work was supported by the FFL18-0277 grant from the Swedish Foundation for Strategic Research.}}

\maketitle

\begin{abstract}

The existence of unknown interference is a prevalent problem in wireless communication networks. Especially in multi-user multiple-input multiple-output (MIMO) networks, where a large number of user equipments are served on the same time-frequency resources, the outage performance may be dominated by the unknown interference arising from scheduling variations in neighboring cells. In this letter, we propose a Bayesian method for modeling the unknown interference power in the uplink of a cellular network. Numerical results show that our method accurately models the distribution of the unknown interference power and can be effectively used for rate adaptation with guaranteed target outage performance.
\end{abstract}


\section{Introduction}

Multi-user multiple-input multiple-output (MU-MIMO) has a great potential to boost the throughput of wireless networks through spatial multiplexing of multiple user equipments (UEs) on the same time-frequency resources \cite{bjornson2017book}. Moreover, network densification is key to satisfying the ever-growing demand in wireless networks \cite{zaher2022soft}. Though the combination of these two concepts can offer high data rates, the strength of the co-channel interference from neighboring cells can vary rapidly due to scheduling decisions and cause outages. The problem is the massive signaling required to measure the instantaneous interference level; thus, it is usually not done in practice \cite{zhang2019learning}.

Several works have focused on determining the signal-to-interference-plus-noise ratio (SINR) distribution in different MIMO setups.
In \cite{gao1998theoretical}, the cumulative distribution function (CDF) of the SINR is derived with equal power interferers, whereas \cite{ali2010performance} derives the SINR distribution in a Poisson field of interferers, both assuming minimum mean square error (MMSE) combining and perfect CSI. In addition, \cite{li2005distribution,ma2008capacity} approximate the SINR distribution by a Gamma random variable when using zero-forcing (ZF) and MMSE receivers with perfect CSI for the equal power case and uniform interferers, respectively. The recent works \cite{zhang2019learning,zhang2019rate} derive asymptotic approximations of the SINR distribution in a Poisson field of interferers with imperfect CSI by fitting a Gamma distribution to the co-channel interference in the uplink (UL) and downlink (DL), respectively. Further, an asymptotic approximation using an exclusion ball model with maximum ratio (MR) and ZF receivers is provided in \cite{bai2016analyzing}, where the approximation becomes tight only when the number of antennas grows large and is much greater than the number of interfering UEs. In \cite{lim2019distribution}, the SINR distribution under uncorrelated Rayleigh fading, equal power interferers, and perfect CSI is presented for an MMSE MIMO system using regular patterns. A common assumption in these works is knowledge of the statistical properties of the underlying interference for computing the outage probability, which leads to expressions that are only valid for a particular set of assumptions; for example, precoding/combining and interference channel distribution. In addition, numerous exponential family distributions have been proposed in Bayesian radar detection problems as in \cite{jay2003bord,wang2006maximum}, to model the variance of a non-Gaussian spherically invariant random process.

In this letter, inspired by the Bayesian radar detection problems, we propose an alternative method to infer the type of distribution of the unknown interference power arising from UEs in neighboring cells. We make use of observations of the total unknown interference power in a relatively large network to compute the sample mean and sample variance for this distribution that, under some practically feasible assumptions, converge to the true mean and variance, respectively. Further, we provide a tool for UL rate adaptation with guaranteed target outage performance based on our analytical work and validate the accuracy of the proposed method numerically.

\section{System Model}\label{model}

The system consists of $K$ single-antenna UEs and a base station (BS) with a receiver array composed of $N$ antennas. The UEs are divided into three categories: the desired UE, $K_n$ known interferers, and $K_u$ unknown interferers, i.e., $K = K_n + K_u + 1$. A standard block-fading channel model is adopted where the time-varying wideband channels are divided into time-frequency coherence blocks with static and frequency-flat channels in each block. The coherence block is composed of $\tau_c$ symbols and the channels take independent random realizations in each coherence block. The channel between UE $k$ and the receiver array is represented by $\mathbf{h}_{k} \in \mathbb{C}^{N \times 1}$ and modelled by correlated Rayleigh fading as $\mathbf{h}_{k} \sim \mathcal{N}_{\mathbb{C}}(\mathbf{0}, \mathbf{R}_{k})$, where $\mathbf{R}_{k} \in \mathbb{C}^{N \times N}$ denotes the spatial correlation matrix. The normalized trace $\beta_{k} = \frac{1}{N} \hspace{1pt}\textrm{tr}\hspace{-1pt}\left(\mathbf{R}_{k}\right)$ determines the average channel gain from a given antenna at the receiver to UE $k$.

\subsection{Channel Estimation}

A time-division duplex (TDD) protocol is adopted consisting of a pilot transmission phase for channel estimation followed by a data transmission phase. The coherence block of $\tau_c$ symbols is divided into $\tau_p$ symbols for UL pilots and $\tau_u$ symbols for UL data transmission, i.e., $\tau_c = \tau_p + \tau_u$.

In the channel estimation phase, each UE is assigned a $\tau_p$-length pilot from a set of $\tau_p$ mutually orthogonal pilots. Since this is not the main focus of the paper, we consider the case that $ K_n < \tau_p < K$, implying orthogonal pilot assignment for UEs belonging to the same cell, whereas uniform random pilot assignment is utilized for the unknown interferers in neighboring cells. We let $\mathcal{P}_t \subset \{1, \hdots, K\}$ denote the UEs that are assigned to pilot $t$. After correlating the received signal with pilot $t$, the signal $\mathbf{y}_{t}^p \in \mathbb{C}^{N \times 1}$ is computed as
\begin{equation}
\mathbf{y}_{t}^p = \sum_{i \in \mathcal{P}_t}\sqrt{\tau_pp_i}\mathbf{h}_{i} + \mathbf{n}_{t},
\end{equation}
where $p_i\hspace{-1pt}$ is the transmit power of UE $i$ and $\mathbf{n}_{t} \hspace{-0.85pt}\sim \mathcal{N}_{\mathbb{C}}\hspace{-0.5pt}\left(\mathbf{0}, \sigma^2\mathbf{I}_N\hspace{-0.5pt}\right)$ denotes the additive Gaussian noise vector at the receiver. We assume that only the statistics of the UEs' channels belonging to the desired cell are known at the receiver. Hence, we employ an estimator that utilizes only the known channel statistics and has a similar structure to the MMSE estimator in \cite{bjornson2017book}. As such, the channel between a known UE $k$ and the receiver is estimated as
\begin{equation}
\hat{\mathbf{h}}_{k} = \sqrt{\tau_pp_k}\mathbf{R}_{k}\left(\tau_pp_k\mathbf{R}_{k} + \sigma^2\mathbf{I}_N\right)^{\hspace{-1pt}-1}\hspace{-1pt}\mathbf{y}_{t}^p.
\end{equation}

Note that the expression in the inverse represents the covariance matrix of the received pilot signal excluding the unknown interferers, meaning that the above estimate is MMSE optimal in the absence of unknown interferers. Since the channel statistics of the unknown interferers that may share the same pilot as UE $k$ are not available at the serving BS, the serving BS is only capable of utilizing the statistics of UE $k$'s channel for determining the estimate.

\subsection{Uplink Data Transmission}

We assume all UEs transmit their signals on the same time-frequency resources, as is customary in cellular MU-MIMO systems. Hence, the received UL signal is given by
\begin{equation}
    \mathbf{y}^{ul} = \sum_{i = 1}^{K}\sqrt{p_i}\mathbf{h}_is_i + \mathbf{n},
\end{equation}
where $s_i$ denotes the zero-mean signal of UE $i$ and $\mathbf{n} \sim \mathcal{N}_{\mathbb{C}}\!\left(\mathbf{0}, \sigma^2\mathbf{I}_N\right)$ corresponds to the additive noise at the receiver. Employing a linear receive combining scheme, the signal estimate for the desired UE $k$ is computed as
\begin{equation}
    \hat{s}_k = \sqrt{p_k}\mathbf{v}_k^H\mathbf{h}_ks_k + \sum_{\substack{i = 1\\i \neq k}}^{K}\sqrt{p_i}\mathbf{v}_k^H\mathbf{h}_is_i + \tilde{n},
\end{equation}
where $\mathbf{v}_k$ denotes the receive combining vector for UE $k$ and $\tilde{n} = \mathbf{v}_k^H\mathbf{n}$ represents the processed noise. Further, we define $\mathcal{D}_u$ as the set containing the $K_u$ unknown interferers and $\mathcal{D}_n$ as the set of $K_n + 1$ known UEs.

A lower bound on the ergodic UL capacity is obtained by utilizing the standard use-and-then-forget (UatF) capacity bounding method from the MU-MIMO literature \cite{bjornson2017book}. Accordingly, an achievable SE for the desired UE $k$ is given by
\begin{equation} \label{eq:SEk}
\textrm{SE}_k = \frac{\tau_u}{\tau_c}\textrm{log}_2\left(1 + \textrm{SINR}_k\right),
\end{equation}
where
\begin{equation}
\textrm{SINR}_k = \frac{|\textrm{DS}_k|^2}{\textrm{IUI}_k^u + \textrm{IUSI}_k^n + \tilde{\sigma}^2}
\label{SINR}
\end{equation}
and
\begin{align}
    &\textrm{DS}_k = \sqrt{p_k}\mathbb{E}\{\mathbf{v}_k^H\mathbf{h}_k\},\\
    &\textrm{IUI}_k^u = \sum_{i \in \mathcal{D}_u}p_i\mathbb{E}\{|\mathbf{v}_k^H\mathbf{h}_i|^2\},\label{unknownint}\\
    &\textrm{IUSI}_k^n = \sum_{i \in \mathcal{D}_n}p_i\mathbb{E}\{|\mathbf{v}_k^H\mathbf{h}_i|^2\} - p_k|\mathbb{E}\{\mathbf{v}_k^H\mathbf{h}_k\}|^2,\\
    &\tilde{\sigma}^2 = \sigma^2\mathbb{E}\{||\mathbf{v}_k||^2\}.
\end{align}

In the above expression, $\textrm{SINR}_k$ represents the effective SINR, $\textrm{DS}_k$ denotes the desired signal over the deterministic average channel, $\textrm{IUI}_k^u$ represents the inter-user interference power arising from UEs in neighboring cells, whereas $\textrm{IUSI}_k^n$ represents the inter-user interference power from UEs in the desired cell plus the self-interference due to channel uncertainty. $\tilde{\sigma}^2$ denotes the variance of the processed noise. Note that the above expression is applicable along with any combining scheme, however, we will employ MR and partial regularized ZF (RZF) combining in the numerical evaluation. The combining vector for UE $k$ can thus be written as
\begin{equation}
\mathbf{v}_{k} =
\begin{cases}
      \hat{\mathbf{h}}_{k}/||\hat{\mathbf{h}}_{k}||^2 & \textrm{for MR}, \\
      \left(\sum\limits_{i \in \mathcal{D}_n}p_i\hat{\mathbf{h}}_{i}\hat{\mathbf{h}}_{i}^H + \sigma^2\mathbf{I}_N\right)^{-1}p_k\hat{\mathbf{h}}_{k} & \textrm{for RZF}.
\end{cases}
\end{equation}

\section{Unknown Interference Power Distribution}

The main focus of this work is to characterize the distribution of the total unknown interference power coming from UEs in neighboring cells. In practice, the unknown interference power from adjacent cells is not known in the desired cell, because it can change rapidly due to user mobility and scheduling decisions in the neighboring cells.

In a dense network, the number of unknown interferers in the vicinity of the desired cell is expected to be large. The total unknown interference represented by the sum of a large number of independent interference terms converges, by the central limit theorem, to a Gaussian distribution; that is, $\sum\limits_{i \in \mathcal{D}_u}\sqrt{p_i}\mathbf{v}_k^H\mathbf{h}_is_i \xrightarrow[|\mathcal{D}_u| \rightarrow \infty]{d} \mathcal{N}_{\mathbb{C}}\left(0, \sigma_s^2\right)$. Note that the signals from the different interfering sources $s_i$, $i \in \mathcal{D}_u$ are independent of each other and of the channels and noise.

A necessary condition for the total unknown interference above to approach a Gaussian distribution is that there is not a single dominant unknown interferer with much larger channel gain than the other unknown interferers, which is not expected in a practical cellular setup that identifies the right AP-UE associations as the presented in Fig.\,1. In the numerical evaluation, we will show the tightness of the derived distribution for the unknown interference power, which is based on that the unknown interference approaches a Gaussian distribution. We present the result for different numbers of unknown interferers and different desired UE locations.

In the following, we derive the distribution of the unknown variance for a Gaussian random variable and utilize the result to characterize the distribution of the unknown interference power.

\begin{theorem} \label{th:inverse-gamma}
\textit{The marginal posterior distribution of the unknown variance of a random variable} $X \sim \mathcal{N}\left(\mu_{u}, \sigma_u^2\right)$ \textit{with known mean has an Inverse-Gamma distribution with a probability density function (PDF) given by}
\begin{equation}
    f\left(\sigma_u^2|x\right) = \frac{\zeta^{\frac{1}{2}}}{\Gamma\left(\frac{1}{2}\right)} \left(\frac{1}{\sigma_u^2}\right)^{\frac{3}{2}}e^{-\frac{\zeta}{\sigma_u^2}},
    \label{IGpdf}
\end{equation}
\textit{where} $\zeta = \frac{\left(x - \mu_{u}\right)^2}{2}$ \textit{and} $\Gamma\left(\cdot\right)$ \textit{is the gamma function.}
\end{theorem}
\begin{IEEEproof}
The proof is given in Appendix A.
\end{IEEEproof}

In practice, the statistics of the total unknown co-channel interferers' power can be estimated by the desired cell. By observing a sufficiently large sample of the interference powers and feasibly assuming that the statistics of the unknown interferers' power vary slowly over the day compared to the desired UE transmission interval, the sample mean and sample variance of the unknown interference power converge to the true mean and variance of the distribution, respectively \cite{zhang2019learning}. Using the result of Theorem~\ref{th:inverse-gamma}, the unknown interference power distribution is given by the following.

\begin{corollary}
\textit{The unknown interference power has an Inverse-Gamma distribution with a PDF given by}
\begin{align}
    &f_{\textrm{IUI}_k^u}\left(x\right) = \frac{\beta^{\frac{3}{2}}}{\Gamma\left(\alpha\right)} \left(\frac{1}{x}\right)^{\alpha + 1}e^{-\frac{\beta}{x}},\\
    &\alpha = \left(\frac{\mu^2}{v}\right) + 2,\\
    &\beta = \left(\frac{\mu^2}{v} + 1\right) \mu,
    \label{Intpdf}
\end{align}
\textit{where $\alpha$ and $\beta$ denote the shape and scale parameters, respectively, and $\mu$ and $v$ represent the estimated mean and variance for the total unknown interference power.}
\end{corollary}

\section{Outage Performance Analysis}

The interference term $\mathrm{IUI}_k^u$ in \eqref{unknownint} rely on average fading coefficients of UEs in neighboring cells that are unknown to the desired BS, which in turn raises an issue for robust rate adaptation because the achievable SE in \eqref{eq:SEk} cannot be computed. The randomness here is due to the unknown UE locations and corresponding unknown channel statistics. 
In practice, the instantaneous realizations of these interference terms are not measured at the desired BS due to the large signaling overhead that would be required \cite{zhang2019learning} and because scheduling decisions are made simultaneously in all cells. In this section, we utilize our derived analytical results to derive the outage probability and corresponding $\epsilon$-outage SE performance for the uplink system described in Section \ref{model}.
The CDF for the SINR expression in \eqref{SINR} is given by
\begin{equation}
\begin{split}
    \textrm{Pr}\left[\textrm{SINR}_k \leq T\right] &= \textrm{Pr}\left[\frac{|\textrm{DS}_k|^2}{\textrm{IUI}_k^u + \textrm{IUSI}_k^n + \tilde{\sigma}^2} \leq T\right]\\
    &= \textrm{Pr}\left[\textrm{IUI}_k^u \geq \frac{|\textrm{DS}_k|^2}{T} - \textrm{IUSI}_k^n - \Tilde{\sigma}^2\right]\\
    &= 1 - F_{\textrm{IUI}_k^u}\left(\frac{|\textrm{DS}_k|^2}{T} - \textrm{IUSI}_k^n - \Tilde{\sigma}^2\right),
\end{split}
\end{equation}
where $F_{\textrm{IUI}_k^u}\left(\cdot\right)$ corresponds to the Inverse-Gamma CDF of the unknown interference power $\textrm{IUI}_k^u$. Further, given a target outage probability $\epsilon$, the $\epsilon$-outage SE for the desired UE $k$ is given by
\begin{equation}
    R_k\left(\epsilon\right) = \frac{\tau_u}{\tau_c}\textrm{log}_2\left(1 +T_k\left(\epsilon\right)\right)
\end{equation}
with $T_k\left(\epsilon\right)$ given by
\begin{equation}
    T_k\left(\epsilon\right) = \frac{|\textrm{DS}_k|^2}{F_{\textrm{IUI}_k^u}^{-1}\left(1 - \epsilon\right) + \textrm{IUSI}_k^n + \tilde{\sigma}^2}.
\end{equation}
Note that all terms, other than the unknown interference power, are assumed to be quasi-static having a fixed known value at the desired BS for a given transmission interval. Moreover, we stress that the outage framework proposed in this section is managing the unknown interference statistics, while traditional outage analysis considers the fundamentally different scenario of having unknown fading realizations for the desired signal.

The proposed method for robust rate allocation can be summarized in the following four-step procedure.

\begin{itemize}    
    \item \textbf{Step\,\,1:} Measure the total unknown interference power $\textrm{IUI}_k^u$ at the serving BS for a certain number of samples. For example, this can be done at slots where the known UEs are silent (not scheduled by the serving BS).
    
    \item \textbf{Step\,\,2:} Compute the sample mean $\mu$ and sample variance $v$ for $\textrm{IUI}_k^u$ based on the samples obtained in Step 1.
    
    \item \textbf{Step\,\,3:} Compute the parameters $\alpha$ and $\beta$ according to (14) and (15), respectively.
    
    \item \textbf{Step\,\,4:} Compute the achievable SE for a given target outage probability $R_k\left(\epsilon\right)$ according to (17) and (18).
\end{itemize}

\section{Numerical Evaluation}

\begin{figure}
\centering
\setlength{\abovecaptionskip}{0.2cm plus 0pt minus 0pt}
\includegraphics[scale=0.49]{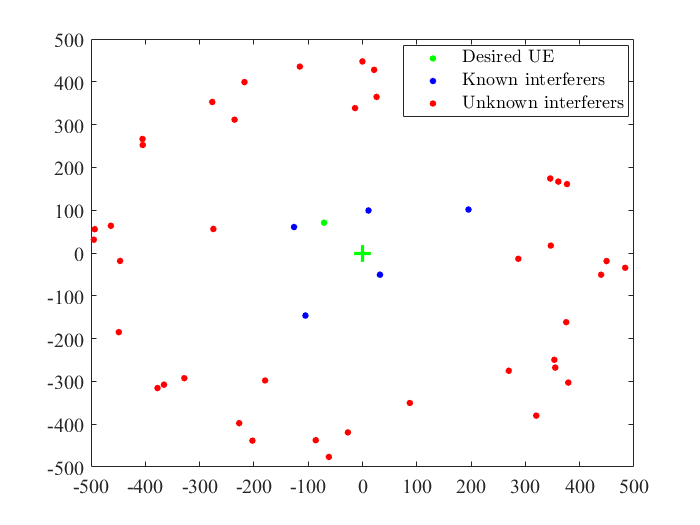}
\caption{Site map and UE distribution with $K_u = 40$ unknown interferers. Serving BS is located at the origin.}
\label{scenario}
\end{figure}

In this section, we corroborate the derived analytical results by Monte Carlo simulations and evaluate the UL SE that can be achieved while satisfying different target outage probabilities. We consider a half-wavelength spaced uniform linear array of $N = 16$ antennas at the desired BS located at the origin. The locations of the desired UE and known interferers are assumed fixed throughout the simulations since we focus on the SINR variations caused by the inter-cell interference that is unknown at the serving BS. We employ two different simulations; the first considers a distance of $r = 100$\,m whereas the second considers $r = 200$\,m between the desired UE and serving BS. In both cases, $K_n = 5$ known interferers are dropped, each on a circle of radius $\{60, 100, 140, 180, 220\}$\,m and having a fixed angle to the serving BS that is chosen in a uniformly random manner. In each simulation instance, $K_u$ unknown interferers are randomly and uniformly dropped on a disk in the radius range of $[250, 500]$\,m. The network simulation parameters are summarized in Table \ref{params}. We utilize the 3GPP Urban Microcell model for generating large-scale fading coefficients with correlated shadowing among the UEs as given in \cite[Eq.~(37)]{bjornson2019making}. 
Fig. \ref{scenario} depicts the simulated setup with a distance of $r = 100$\,m between the desired UE and serving BS, and fixed known interferers' locations. The figure shows a single random realization for the locations of the unknown interferers that are assumed to be associated with BSs in neighboring cells.

\begin{table}
\begin{center}
\caption{Network simulation parameters.}
\begin{tabular}{ |c|c| }
\hline
Bandwidth & $20$\,MHz \\
Number of BS antennas & $N = 16$ \\
Pathloss exponent & $\alpha = 3.76$ \\
UL transmit power & $p_i = 100$\,mW \\
UL noise power & $-94$\,dBm \\
Coherence block length & $\tau_c = 200$ \\
Pilot sequence length & $\tau_p = 10$ \\
\hline
\end{tabular}
\label{params}
\end{center}
\vspace{-1.19em}
\end{table}

\begin{figure}
\centering
\setlength{\abovecaptionskip}{0.33cm plus 0pt minus 0pt}
\includegraphics[scale=0.49]{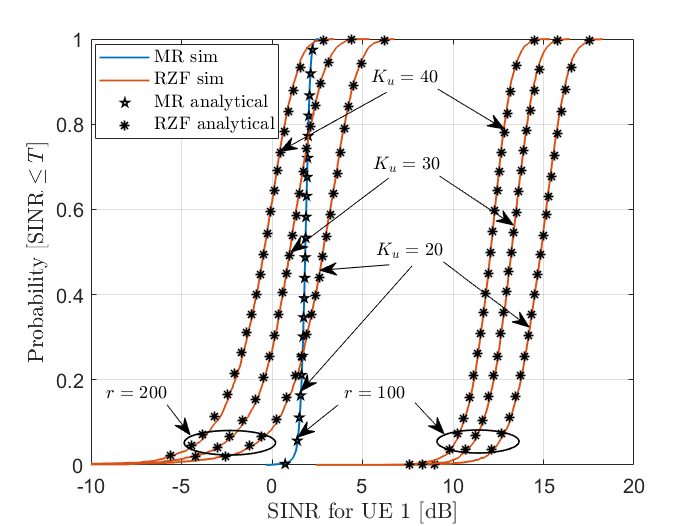}
\caption{Outage probability for different numbers of unknown interferers and different distances between desired UE and BS.}
\label{combined_SINR}
\end{figure}

Fig. \ref{combined_SINR} shows the CDF of the SINR at the desired UE with both the analytical model and Monte Carlo simulation results for different numbers of unknown interferers and different distances between the desired UE and serving BS, denoted by $r$. It is clear that the analytical performance matches tightly with the exact numerical evaluation for all simulated scenarios. The SEs achieved with MR combining are about the same for different numbers of unknown interferers in neighboring cells, thus, we only show the case of $K_u = 20$ unknown interferers. The reason is that MR ignores interference and then the denominator of the SINR is dominated by interference from the known co-channel interferers in the desired cell. Alternatively, when employing RZF, the residual known interference is seen to be weaker than the unknown interference, resulting in CDF curves with larger variations. Further, a noticeable change in SINR is observed with changing the number of unknown interferers, showing that the unknown interference may significantly impact the outage performance.

\begin{figure}
\centering
\setlength{\abovecaptionskip}{0.33cm plus 0pt minus 0pt}
\includegraphics[scale=0.49]{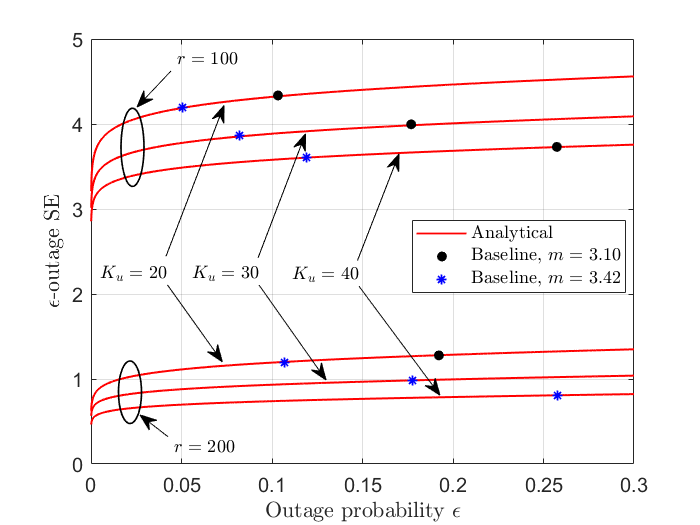}
\caption{$\epsilon$-outage SE with RZF.}
\label{combined_outage}
\end{figure}

To demonstrate the effectiveness of our analytical result, Fig. \ref{combined_outage} plots the $\epsilon$-outage SE with RZF for different numbers of unknown interferers and different distances between the desired UE and serving BS. A baseline scheme, based on the classic idea of having a fixed fade margin is shown for comparison. The baseline utilizes no knowledge of the unknown interference for rate adaptation but instead divides the effective SINR of the desired UE in \eqref{SINR}, excluding the term $\mathrm{IUI}_k^u$, by a fixed margin $m$ to compensate for the unknown interference from UEs in neighboring cells. Precisely, the baseline chooses the SE as
\begin{align}
&\textrm{SINR}_k^{\textrm{bl}} = \frac{|\textrm{DS}_k|^2}{ \textrm{IUSI}_k^n + \tilde{\sigma}^2}\\
&\textrm{SE}_k^{\textrm{bl}} = \frac{\tau_u}{\tau_c}\textrm{log}_2\left(1 + \frac{\textrm{SINR}_k^{\textrm{bl}}}{m}\right).
\end{align}

Note that for $K_u \in \{30, 40\}$, $r = 200$\,m, and $m = 3.10$, the resulting outage probability $\epsilon > 0.3$; hence, it is not shown in the figure. For the baseline scheme, it can be seen that the resulting outage probability varies significantly with the number of unknown interferers $K_u$ and the distance $r$ between the desired UE and BS. This means that it is difficult in practice to determine the right UL transmission rate in order to maintain the same resulting outage probability for different situations. On the other hand, our proposed analytical model is able to determine the correct SE that results in any target predefined outage probability; thus, providing an effective tool for UL rate adaptation with guaranteed target outage performance.

\section{Conclusions}

In this letter, the prevalent problem of unknown UL interference in an MU-MIMO wireless network has been addressed. We derive an analytical model for the unknown interference power distribution and show the accuracy of the analysis numerically. The model is useful for robust rate allocation to handle outages caused by rapidly changing inter-cell interference. A baseline scheme that utilizes a fixed margin to compensate for the unknown interference power has been presented for comparison. The results show that with such a scheme, a minor change in the selected UL SE for transmission or a change in the simulated scenario leads to a significant change in the resulting outage performance. In contrast, our framework is capable of always choosing the correct SE that offers a specific predetermined target outage probability. 

\section*{Appendix A}

The PDF of $X$ given $\sigma_u^{2}$ follows a Gaussian distribution and is given by
\begin{equation}
    f\left(x|\sigma_u^2\right) = \frac{1}{\sqrt{2\pi\sigma_u^2}}e^{-\frac{\left(x - \mu_{u}\right)^2}{2\sigma_u^2}}.
\end{equation}

We utilize an uninformative prior $g\left(\sigma_u^{2}\right) = 1/\sigma_u^2$, i.e., assuming no prior knowledge of the variance distribution. Using Bayes formula, the posterior distribution of $\sigma_u^{2}$ is computed as

\begin{equation}
\begin{split}
    f\left(\sigma_u^2|x\right) &= \frac{f\left(x|\sigma_u^{2}\right)g\left(\sigma_u^2\right)}{f\left(x\right)}\\
    &= \frac{1}{\left(2\pi\right)^{\frac{1}{2}}f\left(x\right)}\left(\sigma_u^2\right)^{-\frac{3}{2}}e^{-\left(\frac{1}{\sigma_u^2}\frac{\left(x - \mu_{u}\right)^2}{2}\right)}.
    \label{fsigma}
\end{split}
\end{equation}

Recognizing that $f\left(x|\sigma_u^{2}\right)g\left(\sigma_u^2\right)$ has a similar structure to the gamma function, integrating $\sigma_u^{2}$ out to get $f\left(x\right)$ leads to
\begin{equation}
\begin{split}
    f\left(x\right) &= \int_{0}^{\infty}\frac{1}{\left(2\pi\right)^{\frac{1}{2}}}\left(\sigma_u^{2}\right)^{-\frac{3}{2}}e^{-\left(\frac{1}{\sigma_u^2}\frac{\left(x - \mu_{u}\right)^2}{2}\right)}d\sigma_u^{2}\\
    &\overset{\left(a\right)}{=} \frac{1}{\left(2\pi\right)^{\frac{1}{2}}}\int_{0}^{\infty}\left(t\right)^{-\frac{1}{2}}e^{-\left(t\frac{\left(x - \mu_{u}\right)^2}{2}\right)}dt\\
    &= \frac{1}{\left(2\pi\right)^{\frac{1}{2}}}\Gamma\left(\frac{1}{2}\right)\left(\frac{\left(x - \mu_{u}\right)^2}{2}\right)^{-\frac{1}{2}},
    \label{fx}
\end{split}
\end{equation}
where we have introduced the variable $t = 1/\sigma_u^2$ in $\left(a\right)$. Plugging \eqref{fx} into \eqref{fsigma} and letting $\zeta = \frac{\left(x - \mu_{u}\right)^2}{2}$, we get
\begin{equation}
    f\left(\sigma_u^2|x\right) = \frac{\zeta^{\frac{1}{2}}}{\Gamma\left(\frac{1}{2}\right)}\left(\frac{1}{\sigma_u^{2}}\right)^{\frac{3}{2}}e^{-\frac{\zeta}{\sigma_u^{2}}},
\end{equation}
which corresponds to an Inverse-Gamma distribution with parameters $1/2$ and $\zeta$.

\section*{References}
\renewcommand{\refname}{ \vspace{-\baselineskip}\vspace{-1.1mm} }
\bibliographystyle{ieeetr}
\bibliography{papercites}

\end{document}